# STUDY OF THE RADIATION HARDNESS OF VCSEL AND PIN ARRAYS

K.K. GAN, W. FERNANDO, H.P. KAGAN, R.D. KASS, A. LAW,

A. RAU, D.S. SMITH†

*Department of Physics, The Ohio State University, Columbus, OH 43210, USA*

M.R.M. LEBBAI, P.L. SKUBIC

*Department of Physics and Astronomy, University of Oklahoma, Norman, OK 73019, USA*

B. ABI, F. RIZARDINOVA

*Department of Physics, Oklahoma State University, Stillwater, OK 74078, USA*

The silicon trackers of the ATLAS experiment at LHC (CERN) use optical links for data transmission. VCSEL arrays operating at 850 nm are used to transmit optical signals while PIN arrays are used to convert the optical signals into electrical signals. We investigate the feasibility of using the devices at the Super LHC (SLHC). We irradiated VCSEL and GaAs PIN arrays from three vendors and silicon PIN arrays from one vendor. All arrays can be operated up to the SLHC dosage except the GaAs PIN arrays which have very low responsivities after irradiation and hence are probably not suitable for the SLHC application.

## 1. Introduction

Optical links are now widely used in high energy physics experiments for data transmission. The links substantially reduce the volume of metallic signal cables freeing up valuable detector space. In addition, the fibers eliminate the cross talk between metallic cables and electrical ground loops between the front-end electronics and the data acquisition system. The wide bandwidth of opto-electronics is well suited for multiplexing many input channels and allows for introduction of error checking and error recovery transmission protocols. These features are especially important in experiments where radiation can induce Single Event Effects (SEE) in the digital electronics. The silicon trackers of the ATLAS experiment at the Large Hadron Collider (LHC) use VCSELs to generate the optical signals at 850 nm and PIN diodes to convert the signals back into electrical signals for further processing. The devices have been proven to be radiation-hard for operation at the LHC.

† Work partially supported by the U.S. Department of Energy under contract No. DE-FG-02-91ER-40690.

The LHC will start operation in 2008. However, an upgrade of the collider, Super LHC, is already being planned for 2015. The SLHC is designed to increase the luminosity of the LHC by a factor of ten to $10^{35}$ $cm^{-2}s^{-1}$. Accordingly, the radiation level at the detector is expected to increase by a similar factor. In this paper, we present a study of the radiation-hardness of the VCSEL and PIN arrays for the SLHC application.

## 2. SLHC Fluences and Test Setup

We use the Non Ionizing Energy Loss (NIEL) scaling hypothesis to estimate the SLHC fluences [1-3] at the present optical link location (PP0) of the pixel detector of the ATLAS experiment. The estimate is based on the assumption that the main radiation effect is bulk damage in the VCSEL and PIN with the displacement of atoms. After five years of operation at the SLHC (3,000 $fb^{-1}$), we expect the silicon component (PIN) to be exposed to a maximum total fluence of 1.5 x $10^{15}$ 1-MeV $n_{eq}/cm^2$ [4]. The corresponding fluence for a GaAs component (VCSEL) is 8.2 x $10^{15}$ 1-MeV $n_{eq}/cm^2$. We study the response of the optical link to a high dose of 24 GeV protons. The expected equivalent fluences at LHC are 2.6 and 1.6 x $10^{15}$ $p/cm^2$, respectively. For simplicity, we present the results from the irradiations with dosage expressed in Mrad using the conversion factor, 1 Mrad = 3.75 x $10^{13}$ $p/cm^2$ for silicon and 4.57 x $10^{13}$ $p/cm^2$ for GaAs. The expected dosages are therefore 69 and 34 Mrad, respectively.

We irradiated optical modules instrumented with one silicon PIN and two GaAs VCSEL arrays from various vendors using 24 GeV protons at the T7 facility of CERN. The PIN and VCSEL arrays coupled to radiation-hard ASICs produced for the current pixel optical link [5], the DORIC (Digital Opto Receiver Integrated Circuit) and VDC (VCSEL Driver Chip). Furthermore, the opto-boards were mounted on a shuttle system which enabled us to easily move in and out of the beam for annealing of the VCSEL arrays. The test system monitored various parameters of the opto-boards throughout the irradiation. In 2007, we also irradiated GaAs PIN arrays from three vendors. These devices were biased during the irradiation but not monitored due to space constraint.

## 3. Radiation-Hardness of VCSEL and PIN Arrays

We characterized the LIV (Light-Current-Voltage) curves of the VCSEL arrays before the irradiation. In 2006, we irradiated the arrays fabricated by three vendors, Optowell, Advanced Optical Components (AOC, 2.5 Gb/s), and ULM Photonics (two varieties, 5 and 10 Gb/s) [6]. All arrays produced large optical power, in excess of 1 mW for the VCSEL current of 7 mA, the rated maximum current of the ULM 10 Gb/s array. This latter array also required higher voltage, ~ 2.3 V, to produce this current. The 5 Gb/s array required somewhat lower voltage to produce this current and the arrays from the AOC and Optowell, required significantly lower voltage. The latter arrays are therefore more suitable for operation at the SLHC because we expect to fabricate the driver and receiver chips using the 0.13 μm process with a thick oxide option which has a maximum operating voltage of 2.5 V. Given that it requires ~ 0.2 V to operate the transistors in the driver chip, the maximum drive current in the ULM arrays is therefore ~ 7 mA. This implies a lower optical power and less efficient annealing of arrays with radiation damage. In 2007, the AOC 2.5 Gb/s arrays were replace by two new devices operating with higher speed, 5 and 10 Gb/s. Figure 1 shows the LIV curves of the these arrays. These arrays produced large optical power and the required forward voltage to produce 7 mA is significantly less than 2.5 V and hence is suitable for operation at the SLHC.

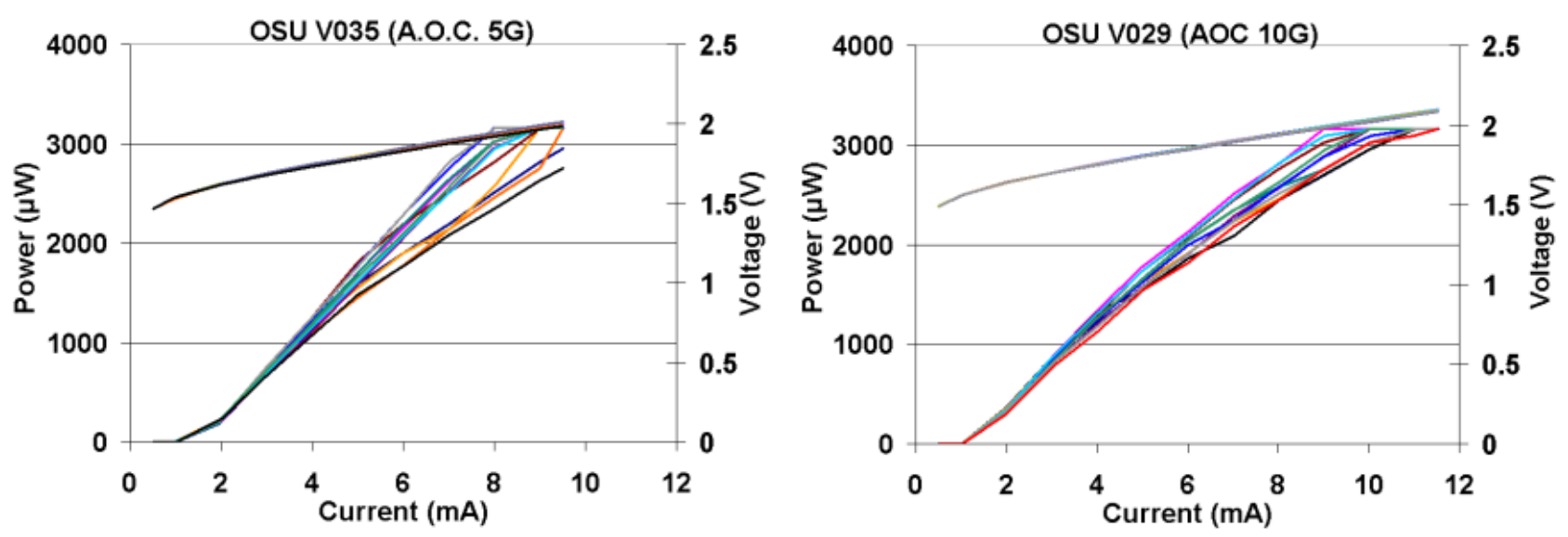

Figure 1. LIV curves of the 5 and 10 Gb/s VCSEL arrays by AOC before irradiation.

The test system monitored the optical power of the VCSEL arrays vs. dosage. In the 2006 irradiation, we found that all arrays continued to produce good optical power up to the SLHC dosage but the degradation with dosage was quite drastic [6]. We believe that the arrays would have performed better should we used a less intense beam and allowed more time for annealing. This is the program we followed in the 2007 irradiation. Figure 2 shows the optical power vs. dosage for the various arrays. The power decreased during the irradiation as expected. We annealed the arrays by moving the opto-boards out of the beam and passing the maximum allowable current (~ 10 mA per channel) through the arrays for several hours each day. The optical power increased during the annealing. Unfortunately, there was insufficient time for a complete annealing. However, all devices continued to produce good optical power up to the SLHC dosage of 34 Mrad, except ULM 5 Gb/s which was least radiation-hard, consistent with the observation of 2006.

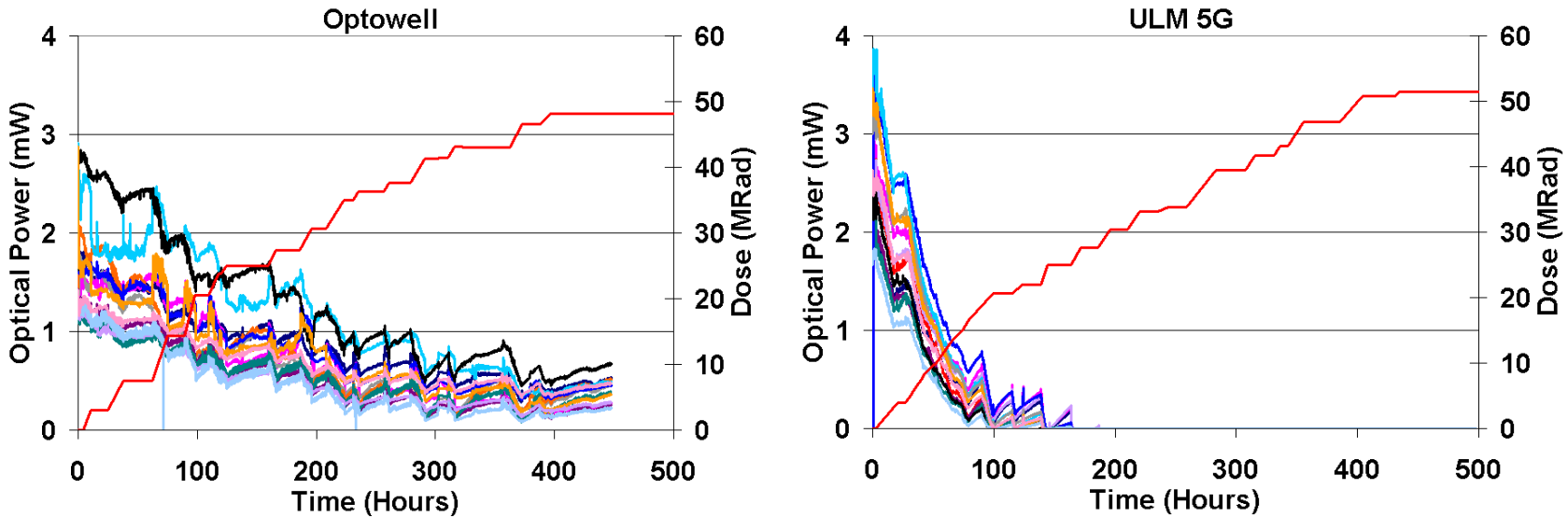

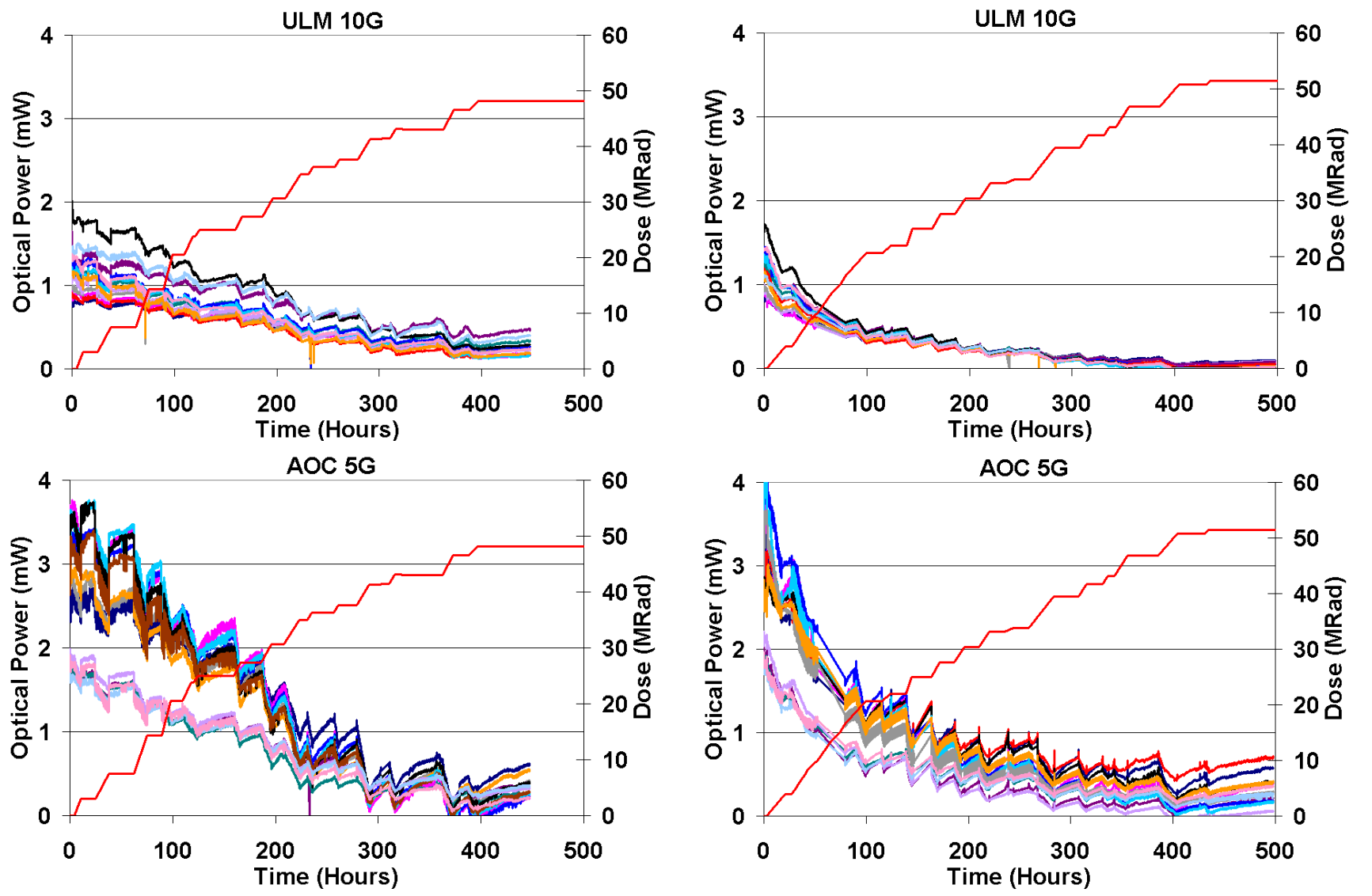
ULM 10G
ULM 10G
AOC 5G
AOC 5G
Optical Power (mW)
Dose (MRad)
Time (Hours)
0
1
2
3
4
0
10
20
30
40
50
60
0
100
200
300
400
500

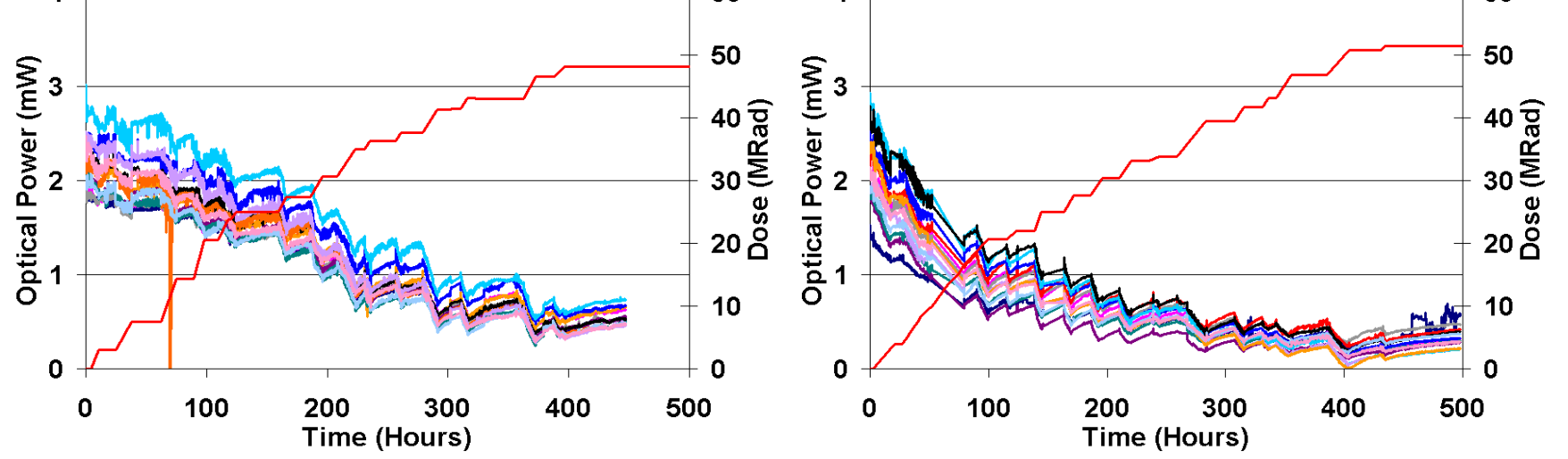
AOC 10G
AOC 10G
Optical Power (mW)
Dose (MRad)
Time (Hours)
0
1
2
3
4
0
10
20
30
40
50
60
0
100
200
300
400
500

Figure 2: Optical power as a function of time (dosage) for various VCSEL arrays that transmitted data to the control room. The power decreased during the irradiation but increased during the annealing as expected.

We irradiated silicon PIN arrays by Truelight in 2006 and found that the responsivities decreased by 65% after the radiation, which is acceptable for the SLHC application. In 2007, we irradiated GaAs VCSEL arrays by AOC, Optowell, and ULM and found that the responsivities decreased by by ~ 90% which is not acceptable for the SLHC application.

## 4. Summary

We have irradiated VCSEL and PIN arrays from various vendors to the SLHC dosage. The GaAs VCSEL arrays from three vendors have been found to have the radiation hardness suitable for the SLHC operation. The responsivities of silicon PIN arrays decrease by ~ 65% while that for GaAs devices decrease by ~ 90% which are not suitable for the SLHC application.

## Acknowledgments

The authors are indebted to M. Glaser in the use of the T7 irradiation facility at CERN.